\documentclass{mem}
\usepackage{natbib}\usepackage{txfonts}
\usepackage{graphicx}
\bibpunct{(}{)}{;}{a}{}{,}

\idline{76}{1}
\begin{document}
\def\teff{$T\rm_{eff }$}
\def\kms{$\mathrm {km s}^{-1}$}
\def\dss{$\delta$~Scuti star}

\title{
Amplitude Ratios as Mode Characterizors in $\delta$ Scuti stars
}

   \subtitle{}

\author{
J.\,R.\,Rasmussen\inst{1} 
\and T.\,H.\,Dall\inst{2}
\and S.\,Frandsen\inst{1}
          }

  \offprints{T. H. Dall}

\institute{
Department of Physics and Astronomy, Ny Munkegade Bldg 520, Aarhus University, 8000 Aarhus C, Denmark
\and
European Southern Observatory, Casilla 19001, Santiago 19, Chile\\
\email{tdall@eso.org}
}

\authorrunning{Rasmussen, Dall \& Frandsen}

\titlerunning{Amplitude Ratios as Mode Characterizors}

\abstract{
Seismology of \dss s
holds great potentials for testing theories of
stellar structure and evolution. 
The ratio of mode amplitudes in light and in equivalent width of spectral
lines can be used for mode identification. However, the amplitude ratios
(AR) predicted from theory are usually inconsistent with observations.
We here present the first results from a campaign aimed at calibrating
observationally the absolute values of the AR.
}
\maketitle{}

\section{Introduction}

The \dss s are A--F non-radial pulsators located near the base of the classical instability strip.
Despite huge efforts it has turned out to be very difficult to
match the observed mode frequencies with calculated ones, due to the large set of poissible
excited modes and the small subset of possible modes actually excited.
\begin{table*}[t!]
\label{tab:xcae}
\caption{Amplitudes measured for X Cae and AI Vel. Uncertainties on amplitudes are 2-3mmag for photometry and 5-8\emph{promille}
for spectroscopy.}
\begin{tabular}{llrrrr} \hline
            & mode frequency & A($B$)  &   A($V$)   &  A($\Lambda_{\mathrm{H}\alpha}$) & R$\alpha$  \\
            &   ~[cd$^{-1}$]    & [mmag]  &  [mmag]    &      [promille]                  &   \\ \hline
{\bf X Cae} &  $f_1 = 7.394$              &   48.4         &   37.9            &   93.9   &  2.5         \\
            &  $f_2 = 6.036$              &   13.5         &   9.9             &   30.8   &  3.1        \\ \hline
{\bf AI Vel} & $f_1 = 8.9627$             & 212.9          & 163.9            &  219.4    &  1.34    \\ 
             & $f_2 = 11.5998$            & 193.2          & 149.2            &  222.7    &  1.49    \\ \hline
\end{tabular}
\end{table*}

Photometric observations of amplitude ratios and phase differences have been used to try to
resolve this mode identification problem \citep[e.g.][]{garrido2000,paparo+sterken2000}, but
the results obtained so far are not unambiguous.  A possible better discriminator is the
ratio of mode amplitudes in light and in equivalent width of spectral
lines, denoted $R\alpha$ for the H$\alpha$ line versus $V$ band photometry.
This ratio has been explored by several
authors \citep[e.g.][]{viskum+1998,frandsen2000,dall+2002,dall+2003}.
However, the interpretation is not straightforward: The amplitude ratios
predicted from theory are usually inconsistent with observations as shown
by \citet{dall+frandsen2002}.

For these reasons we have undertaken a large campaign to establish an empirical calibration of
the amplitude ratios, observing stars with well characterized 
pulsation modes to map the relationships and dependencies on stellar parameters like rotation,
evolutionary stage, spectral type, etc.   The campaign has been concluded, and we present here
a progress report with initial results for two stars;
AI~Vel (HD\,69213, HIP\,40330) and X~Cae (HD\,32846, $\gamma^2$\,Cae, HIP\,23596).

\section{Observations and Data Analysis}
The project sample consists of 16 \dss s, chosen to cover a broad range of amplitudes
and to include both HADS and low-amplitude pulsators.

Photometry was collected with the 24-inch telescope at Siding Spring Observatory, Australia, while the
spectroscopy was obtained with the Danish 1.54m telescope and DFOSC at the European
Southern Observatorys La Silla site in Chile.

X Cae is a well studied \dss\ \citep{mantegazza+poretti1992,mantegazza+poretti1996,mantegazza+2000} showing a wealth of excited modes.
\citet{mantegazza+2000} found 17 frequencies, of which a few showed amplitude variations between
observing seasons.  AI Vel is a well known HADS, that has been studied since the works of \citet{walraven1952,walraven1955}. 
\citet{walraven+1992} found that neither of the two main modes had changed amplitude in 40 years.

Our data are too sparsely sampled to allow anything but the strongest modes to
be confirmed in either star (Table~\ref{tab:xcae}).
Our X~Cae amplitudes of $f_1$ and $f_2$ agree very well with the findings of Mantegazza et al., supporting
their conclusion that these two modes do not change amplitude.

\section{Interpreting the Amplitude Ratios}
Balmer lines have sensitivity similar to radial velocity measurements because of the strong limb darkening 
in these lines, while photometry has very weak center-to-limb variation. Thus the equivalent width of Balmer 
lines and the photometry show different response to spatial variations across the stellar disk (i.e. to the $\ell$ value),
reflected in the amplitude ratio.  Intuitively, we expect radial modes to have the lowest values of $R\alpha$
with higher-order modes having progressively higher ratios.  Large variations from star to star may be expected due
to the wide range of rotational velocities encountered in \dss s.

For BN Cnc \citep{dall+2002} and FG~Vir \citet{viskum+1998} we found amplitude ratios for the radial modes
around $R \sim 0.5$. Looking at the results for X~Cae and AI~Vel we find that the strongest modes in these stars
may actually be non-radial modes, contrary to previous findings --- however, the analysis has not yet been completed.



\bibliographystyle{bibtex/aa}

\bibliography{tdall}

\begin{thebibliography}{13}
\expandafter\ifx\csname natexlab\endcsname\relax\def\natexlab#1{#1}\fi

\bibitem[{{Dall} \& {Frandsen}(2002)}]{dall+frandsen2002}
{Dall}, T.~H. \& {Frandsen}, S. 2002, \aap, 386, 964

\bibitem[{{Dall} {et~al.}(2002){Dall}, {Frandsen}, {Lehmann}, {Anupama},
  {Kambe}, {Handler}, {Kawanomoto}, {Watanabe}, {Fukata}, {Nagae}, \&
  {Horner}}]{dall+2002}
{Dall}, T.~H., {Frandsen}, S., {Lehmann}, H., {et~al.} 2002, \aap, 385, 921

\bibitem[{{Dall} {et~al.}(2003){Dall}, {Handler}, {Moalusi}, \&
  {Frandsen}}]{dall+2003}
{Dall}, T.~H., {Handler}, G., {Moalusi}, M.~B., \& {Frandsen}, S. 2003, \aap,
  410, 983

\bibitem[{{Frandsen}(2000)}]{frandsen2000}
{Frandsen}, S. 2000, in ASP Conf. Ser. 210: Delta Scuti and Related Stars,
  428--+

\bibitem[{{Garrido}(2000)}]{garrido2000}
{Garrido}, R. 2000, in ASP Conf. Ser. 210: Delta Scuti and Related Stars, 67--+

\bibitem[{{Mantegazza} \& {Poretti}(1992)}]{mantegazza+poretti1992}
{Mantegazza}, L. \& {Poretti}, E. 1992, \aap, 255, 153

\bibitem[{{Mantegazza} \& {Poretti}(1996)}]{mantegazza+poretti1996}
{Mantegazza}, L. \& {Poretti}, E. 1996, \aap, 312, 855

\bibitem[{{Mantegazza} {et~al.}(2000){Mantegazza}, {Zerbi}, \&
  {Sacchi}}]{mantegazza+2000}
{Mantegazza}, L., {Zerbi}, F.~M., \& {Sacchi}, A. 2000, \aap, 354, 112

\bibitem[{{Papar{\' o}} \& {Sterken}(2000)}]{paparo+sterken2000}
{Papar{\' o}}, M. \& {Sterken}, C. 2000, \aap, 362, 245

\bibitem[{{Viskum} {et~al.}(1998){Viskum}, {Kjeldsen}, {Bedding}, {dall},
  {Baldry}, {Bruntt}, \& {Frandsen}}]{viskum+1998}
{Viskum}, M., {Kjeldsen}, H., {Bedding}, T.~R., {et~al.} 1998, \aap, 335, 549

\bibitem[{{Walraven}(1952)}]{walraven1952}
{Walraven}, T. 1952, \bain, 11, 421

\bibitem[{{Walraven}(1955)}]{walraven1955}
{Walraven}, T. 1955, \bain, 12, 223

\bibitem[{{Walraven} {et~al.}(1992){Walraven}, {Walraven}, \&
  {Balona}}]{walraven+1992}
{Walraven}, T., {Walraven}, J., \& {Balona}, L.~A. 1992, \mnras, 254, 59

\end{thebibliography}

\end{document}